\documentclass[prd,tightenlines,nofootinbib,superscriptaddress]{revtex4}
\usepackage[utf8x]{inputenc}
\usepackage{amsmath} 
\usepackage{amsfonts,hyperref}
\usepackage{slashed}
\usepackage[dvips]{graphicx}
\usepackage{color}
\usepackage{bbm}
\allowdisplaybreaks

\usepackage{color}

\begin{document}
 \author{{\bf Damien Bégué}\email{damienb@kth.se}}
  \affiliation{Max Planck Institute for extraterrestrial Physics, Giessenbachstrasse 1, 85748 Garching, Deutschland
}
 \author{{\bf Clément Stahl}}\email{clement.stahl@pucv.cl}
 \affiliation{Instituto de F\'{\i}sica, Pontificia Universidad Cat\'{o}lica de Valpara\'{\i}so, Casilla 4950, Valpara\'{\i}so, Chile}
 \author{{\bf She-Sheng Xue}\email{xue@icra.it}}
 \affiliation{ICRANet, Piazzale della Repubblica 10, 65122 Pescara, Italy}

\date{\today}

\title{A model of interacting dark fluids tested with supernovae and Baryon Acoustic Oscillations data}

\begin{abstract}
We compare supernovae and Baryon Acoustic Oscillations data to the predictions of a cosmological model of interacting dark matter and dark energy. This theoretical model can be derived from the effective field theory of Einstein-Cartan gravity with two scaling exponents $\delta_G$ and $\delta_{\Lambda}$, related to the interaction between dark matter and dark energy. We perform a $\chi^2$ fit to the data to compare and contrast it with the standard $\Lambda$CDM model. We then explore the range of parameter of the model which gives a better $\chi^2$ than the standard cosmological model. All those results lead to tight constraints on the scaling exponents of the model. Our conclusion is that this class of models, provides a decent alternative to the $\Lambda$CDM model.
\end{abstract}

\maketitle

\section{Introduction}
\par The acceleration of the expansion of the Universe has come to a huge surprise to most of the scientific community more than 15 years ago \cite{Riess:1998cb,Perlmutter:1998np}. In modern cosmology, many hypothesis have been proposed to drive the accelerated period of expansion. Among these models is the ubiquitous dark fluid christen dark energy having an equation of state (EoS) corresponding to vacuum energy. This proposal is part of the standard model of cosmology called $\Lambda$CDM which requires the universe to be made of an hypothetical dark energy fluid, in addition to the baryonic matter and dark matter. Since 1999, many independent cosmological observations, such as Cosmic Microwave Background radiation \cite{Ade:2015xua}, have pointed towards the existence of dark energy. However, if its existence relies on a well established observational basis, its nature is still an opaque mystery. Different proposals include, among others, dynamical dark energy (quintessence) \cite{Tsujikawa:2013fta}, dynamical backreaction \cite{Buchert:2007ik}, inhomogeneous cosmologies \cite{Bolejko:2011jc,Stahl:2016vcl} and modified gravity \cite{Joyce:2014kja}. More recently, mainly on phenomenological ground, it has been discovered that allowing an interaction between dark energy and dark matter offers an interesting and attractive alternative to the standard model of cosmology. Such a phenomenological interaction is possible as the nature of the dark matter fluid is also poorly constrained and that general relativity only requires the total energy momentum to be conserved, but not the partial ones.

Interacting dark energy models \cite{Friedman:1991dj,Gradwohl:1992ue,Wetterich:1994bg,Amendola:1999er,Amendola:2003wa,Lee:2006za,Pettorino:2008ez,Valiviita:2008iv,He:2008tn,Gavela:2009cy,Faraoni:2014vra,Salvatelli:2014zta,Xue:2014kna,Abdalla:2014cla,Koivisto:2015qua,Kumar:2016zpg,Kumar:2017dnp} rely on the idea that dark energy and dark matter do not evolve separately but interact with each other non-gravitationally (see also the recent review \cite{Wang:2016lxa} and references therein). Most of the studies on interacting dark fluids focus on the relation to the cosmological data introducing an \textit{ad hoc} coupling between dark matter and dark energy \mbox{\cite{Salvatelli:2014zta,Murgia:2016ccp}}. A classification of those models was given in \mbox{\cite{Koyama:2009gd}}. In this work, we adopt a theoretical model arising directly from the 
quantum field theory of Einstein-Cartan gravity \mbox{\cite{Xue:2014kna,Xue:2015tmw}} and perform the link to a set of observational data: the supernovae Ia (hereafter SNe Ia).

Interacting dark energy models are known to solve the coincidence problem: in the present universe the contributions of dark matter and dark energy to the cosmic budget are of the same order, while their respective evolutions are drastically different throughout the cosmic history. Let $\rho_c^0 = 3 H_0^2 /(8\pi G_0)$ be the critical energy density and $\rho_{\rm m,\Lambda}$ be the energy densities of the dark matter and dark energy fluids. Here $H_0$ is the Hubble constant and $G_0$ is the gravitational constant. In the following, quantities with a subscript or superscript $0$ are measured in the present universe. We define $\Omega_{\rm m}$ and $\Omega_{\Lambda}$ as the ratio between energy densities of dark matter and dark energy and the critical energy density: 
\begin{equation}
\Omega_{ {\rm m},\Lambda} 
\equiv \frac{\rho_{{\rm m},\Lambda}}{\rho_c^0}.
\end{equation}
Knowing that the parameters $\Omega_{\rm m}$ and $\Omega_{\Lambda}$ vary differently with redshift $z$, with $\Omega_{\rm m} \propto (1+z)^3$ and $\Omega_{\Lambda} \propto \Omega_{\Lambda}^0$, the fact that $ \Omega_{\rm m}^0 \sim \Omega_{\Lambda}^0$ is fairly unnatural: we are observing our Universe in a very peculiar epoch. Observe that with these definitions, the Friedmann equation is nothing but a constraint equation between the two fluids, namely $\Omega_{\rm m}^0+\Omega_{\Lambda}^0=1$.

The model considered in this paper relies on the Einstein-Cartan gravitational theory and the universe would be in the scale invariant ultra-violet fixed point of the theory, which we refer as {\it Quantum Field Cosmology} (QFC) model. It has been shown in \cite{Xue:2014kna,Xue:2015tmw} that it is possible from this very general theory to propose an expression for the luminosity distance, and hence compare it to observational data. This paper aims at realizing this possibility and investigating whether this model, as other interacting dark energy models, offers a viable alternative to the $\Lambda$CDM model. Our article shares the same spirit as \cite{Sola:2016jky} or \cite{Novikov:2016hrc,Novikov:2016fzd}: not only it provides a phenomenological interacting dark energy framework, but it also presents theoretical motivations arising from quantum field theory. This article is divided as follows: in section \ref{sec:SFC} and \ref{sec:latea}, we present the basic equations of the QFC model and calculate an expression for the luminosity distance. In section \ref{sec:fit}, we perform a $\chi^2$ fit to the supernovae data. In section \ref{sec:fit}, we include a second data set from Baryon Acoustic Oscillations (BAO) and perform a joint data analysis. Motivated by the encouraging results, we investigate in section \ref{sec:paramstudy} more systematically the parameter space of our model by asking the following question: which range of the parameter space of the QFC model offers a better alternative than the $\Lambda$CDM model from the supernovae data. Finally, we give conclusions and perspectives in section \ref{sec:ccl}.

\section{Quantum Field Cosmology}
\label{sec:SFC}
As one of the fundamental theories for interactions in Nature, the classical Einstein theory of gravity, which plays an essential role in the standard model of modern cosmology ($\Lambda$CDM), should be realized in the scaling-invariant domain of a fixed point of its quantum field theory\footnote{It was suggested by Weinberg \cite{Weinberg} that the quantum field theory of gravity regularized with an ultraviolet (UV) cutoff might have a non-trivial UV-stable fixed point and asymptotic safety, namely the renormalization group (RG) flows are attracted into the UV-stable fixed point with a finite number of physically renormalizable operators for the gravitational field.}. 
It was proposed \cite{Xue:2014kna,Xue:2015tmw} that the present (low-redshift $z<1$) cosmology is realized in the scaling-invariant domain of an ultraviolet-stable fixed point ($\sim G_0$) of the quantum field theory of Einstein gravity\footnote{Instead, the inflationary cosmology could be realized in the scaling-invariant domain of an ultraviolet-unstable fixed point $\tilde{G}_0 \neq G_0$.}, and is described by
\begin{eqnarray}
H^2
&=& H_0^2
\Big[\Omega^0_{\rm m}a^{-3+\delta_{G}} + 
\Omega^0_{_\Lambda}a^{-\delta_{\Lambda}}\Big],
\label{re3}\\
a\frac{dH^2}{da}\! +\! 2H^2 \!&= &\!H^2_0
\Big[2\Omega^0_{_\Lambda}a^{-\delta_{\Lambda}}\!-\!(1\!+\!3\omega_{\rm m})\Omega^0_{\rm m}a^{-3+\delta_{G}}\Big].
\label{e2}
\end{eqnarray}
Here, $a$ is the scale factor, $H = \dot{a}/a$ is the Hubble parameter. Here $p_{\Lambda,{\rm m}}$ are the pressures of the dark fluids. In deriving Equations (\ref{re3}) and (\ref{e2}), motivated by observations, it was assumed that the curvature is null $k=0$, implying $\Omega_\Lambda^0 + \Omega_{\rm m}^0 = 1$. In addition, in the framework of QFC model, both the gravitation constant and $\Omega_\Lambda/\Omega_\Lambda^0$ can vary, following the scaling evolutions $G/G_0\approx a^{\delta_G}$ and $\Omega_\Lambda/\Omega^0_{\Lambda}\approx a^{-\delta_\Lambda}$, where $a = (1+z)^{-1} \gtrsim 1$. In other words, the evolution of the two dark sectors is described by the two critical indexes $\delta_{G}$ and $\delta_\Lambda$. 
The dark energy and matter interact and can be converted from one to another. They obey the generalized Bianchi identity (total energy conservation), 
\begin{eqnarray}
a\frac{d}{da}\left[(G/G_0)(\Omega_{_\Lambda}+\Omega_{\rm m})\right]
&=&-3(G/G_0)(1+\omega_{\rm m})\Omega_{\rm m},
\label{cgeqi20}
\end{eqnarray}
where effective variations of the gravitational coupling constant and of the cosmological constant generalize the standard Bianchi identity. For small redshift, assuming $\delta_\Lambda <\delta_G\ll 1$, Equation (\ref{cgeqi20}) leads to the relation
\begin{equation}
\label{eq:rela}
\delta_{\Lambda} = \left(\frac{\Omega_{\rm m}^0}{\Omega_{\Lambda}^{0}}\right)\delta_G>0.
\end{equation}
In this article, such a quantum field cosmology (QFC) model with theoretical parameters $\Omega_{\rm m}^0$, $\Omega_\Lambda^0$, $\delta_G$ and $\delta_\Lambda$ is compared with the observational cosmology, the case of 
$\delta_{G} =\delta_{\Lambda}=0$ reducing to the $\Lambda$CDM model.

\section{Effective EoS and interaction of dark energy and matter}
\label{sec:latea}
In the $\Lambda$CDM model, the Friedmann equations read:
 \begin{align}
& H^2 =\frac{8 \pi G}{3} (\rho_{\rm m} +\rho_{\Lambda}), \label{eq:fried1} \\
& \frac{\ddot{a}}{a} = -\frac{4 \pi G}{3} (\rho_{\rm m} + \rho_{\Lambda}+3p_{\rm m} + 3 p_{\Lambda}) \label{eq:fried2}
 \end{align}
Equations (\ref{re3}) and (\ref{e2}) in the QFC model can be obtained by phenomenologically introducing a slight deformation of the evolution of the dark sector of our Universe, obtained from the Friedmann equations:
\begin{align}
& \rho_{\rm m} = \rho_{\rm m}^0 a^{-3+\delta_G}, \label{eq:8}\\
& \rho_{\Lambda} = \rho_{\Lambda}^0 a^{-\delta_{\Lambda}}.\label{eq:9}
\end{align}
Using definitions (\ref{eq:8}) and (\ref{eq:9}) in equations (\ref{eq:fried1}) and (\ref{eq:fried2}), one recovers  equations (\ref{re3}) and (\ref{e2}) of the QFC model, given the EoS $p_\Lambda = - \rho_\Lambda$ and $p_{\rm m} = 0$. Therefore, the parameters $\delta_G$ and $\delta_{\Lambda}$ of equations (\ref{eq:8}) and (\ref{eq:9}) correspond to the parameters of the QFC model discussed in section \ref{sec:SFC}. The total conservation of dark energy and matter leads to the constraint 
(\ref{eq:rela}) on $\delta_G$ and $\delta_{\Lambda}$.

Although the parameters $\delta_G$ and $\delta_{\Lambda}$ are motivated by the QFC model, these parameters can also be explored on phenomenological grounds. It leads to another interpretation of parameters $\delta_G$ and $\delta_\Lambda$.

Using the individual conservation of the energy momentum tensors for the dark matter and dark energy sectors respectively
\begin{align}
\dot{\rho}_{\rm m,\Lambda} + 3 H (1+ \omega_{\rm m,\Lambda}) \rho_{\rm m,\Lambda} =0,
\end{align}
and the Friedmann equation (\ref{eq:fried1}), we find that the parameters  $\delta_G$ and $\delta_\Lambda$ in Eq.~(\ref{re3}) can be interpreted as effective modifications of the EoS with
\begin{align}
& \omega_{\rm m} = -\frac{\delta_G}{3} \\
& \omega_{\Lambda} = -1+\frac{\delta_{\Lambda}}{3}.
\end{align}

Besides, assuming the standard EoS ($\omega_{\rm m}=0, \omega_{\Lambda}=-1$), it is also possible to relate the parameters $\delta_G$ and $\delta_{\Lambda}$ to an interaction between dark matter and dark energy by introducing an interaction term $Q$
\begin{align}
&\dot{\rho}_{\rm m} + 3 H \rho_{\rm m} =+Q, \\
& \dot{\rho}_{\Lambda}  =-Q.
\end{align}
In this case, the interaction term $Q$ can be effectively expressed:
\begin{align}
Q = H \delta_G \rho_{\rm m} = H \delta_{\Lambda} \rho_{\Lambda},
\label{eq:Q}
\end{align}
which leads also to the relation (\ref{eq:rela}). Such interaction terms have been shown to alleviate the coincidence problem \cite{Abdalla:2014cla}. We stress that this last interpretation in term of $Q$ of the deformation the dark sector (\ref{eq:8}) (\ref{eq:9}) is valid only for small redshifts ($z \ll 1$), as otherwise the different evolutions in redshift for $\rho_{\rm m}$ and $\rho_{\Lambda}$ invalidate (\ref{eq:Q}). To obtain the general interaction term, one needs to consider the general evolution of the effective gravitational constant in equation (36) of \cite{Xue:2014kna} but this investigation is out of the scope of this paper.
\paragraph*{} In this article, we will treat the parameters $\delta_G$ and $\delta_\Lambda$ in Eq.~(\ref{re3}) as free parameters determined by the observational cosmology. Observe that in order to have a coherent model of dark energy, the constraint $\delta_G \delta_{\Lambda} >0$ must be fulfilled. In this case, the parameters $\delta_G$ and $\delta_{\Lambda}$ can be interpreted as the rate of conversion of dark matter into dark energy and of dark energy into dark matter. A phenomenological investigation shows that increasing $\delta_G$ or decreasing $\delta_{\Lambda}$ induces an acceleration of the expansion of the universe: dark matter is converted into dark energy. Conversely, decreasing $\delta_G$ or increasing $\delta_{\Lambda}$ induces a deceleration of the expansion of the universe: dark energy is converted into dark matter.

The luminosity distance can be obtained
\begin{align}
\label{eq:dl}
 d_L (z) = \frac{c}{H_0} (1+z) \int_1^{\frac{1}{1+z}} \frac{da}{a^2 \sqrt{\Omega^0_{\Lambda} a^{-\delta_{\Lambda}}+\Omega_{\rm m}^0 a^{\delta_G -3} }},
 \end{align}
where the speed of light $c$ is included for clarity. An analytic representation of the integral reads
 \begin{align*}
 d_L(z)= & \frac{2c}{H_0}\frac{1+z}{\Omega_{\rm m}^0 (1-\delta_G)} \times \Bigg[\text{ }_2F_1\bigg(1,\frac{4-\delta_{\Lambda}-2 \delta_G}{2(3-\delta_{\Lambda}-\delta_G)},\frac{7-\delta_{\Lambda}-3 \delta_G}{2(3-2\delta_{\Lambda}-\delta_G	)};-\frac{\Omega^0_{\Lambda}}{\Omega_{\rm m}^0} \bigg) \\
& -\frac{\sqrt{(1+z)^{3-\delta_G}\Omega_{\rm m}^0+(1+z)^{\delta_{\Lambda}}\Omega_{\Lambda}^0}}{(1+z)^{2-\delta_G}} \text{ }_2F_1\bigg(1,\frac{4-\delta_{\Lambda}-2 \delta_G}{2(3-\delta_{\Lambda}-\delta_G)},\frac{7-2\delta_{\Lambda}-3 \delta_G}{2(3-\delta_{\Lambda}-\delta_G	)};-\frac{\Omega^0_{\Lambda}}{\Omega_{\rm m}^0} (1+z)^{-3+\delta_G +\delta_{\Lambda}}\bigg)\Bigg] ,
 \end{align*}
 where $\text{}_2F_1$ is the hypergeometric function defined as: $\text{}_2F_1(a,b,c;z) \equiv \sum_{n=0}^{\infty}\frac{(a)_n (b)_n}{(c)_n} \frac{z^n}{n!}$, with the Pochhammer symbol $(x)_n$ given by: $(x)_n \equiv \frac{\Gamma(x+n)}{\Gamma(x)}$. This equation is derived under the assumption of a flat universe, that is to say $\Omega_{\rm m}^0 + \Omega_{\Lambda}^0 =1$. We now turn to the analysis of the observational data in the light of the QFC model.
 \section{Fit to supernovae data}
 \label{sec:fit}
\begin{figure}[!htb]
     \includegraphics[scale=0.9]{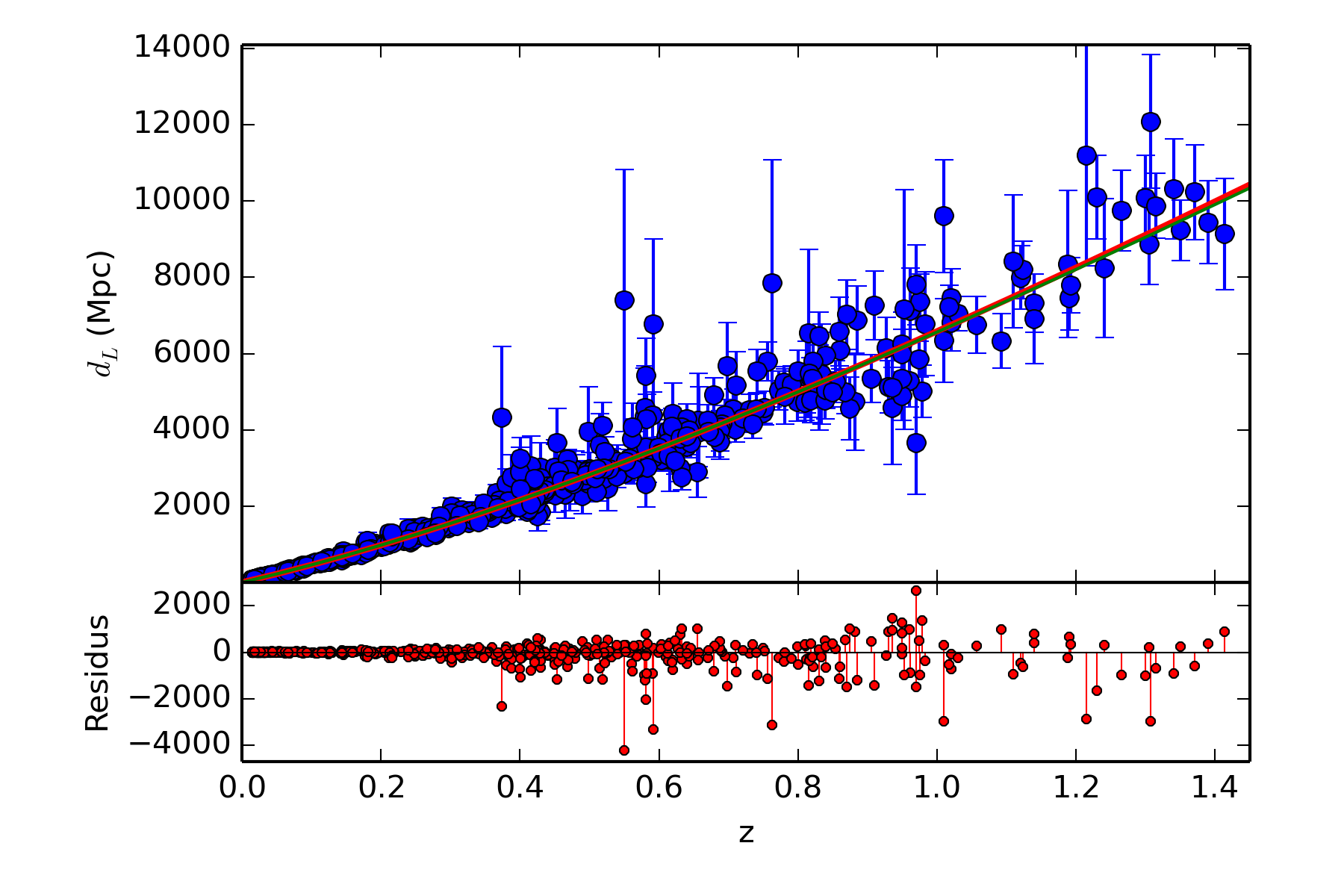}
\caption{Best fitting line for the FLRW model (red, upper curve) and the QFC model (green, lower curve), the quantitative results are given in table \ref{table} (upper panel). Residuals errors of the $\chi^2$ fit of the QFC model (lower panel).}
       \label{fig:1} 
\end{figure}

In this section and in the next one, the relation given by equation (\ref{eq:rela}) is enforced. It gives a constraint between the two new parameters of the QFC model. In section \ref{sec:paramstudy}, this constraint is relaxed in order to study a broader range of the parameter space. We are now in position to perform a $\chi^2$ fit with the Union 2.1 compilation released by the Supernova Cosmology Project \cite{Union}. It is composed of 580 uniformly analyzed SNe Ia. We minimize the reduced $\chi^2$ defined as:
\begin{equation}
\label{eq:chiSN}
\chi^2_{\text{SNe}}(\text{parameters}) \equiv \frac{1}{N_{dof}}\sum_{i=1}^{580} \left[ \frac{d_L(i)-d_L(\text{parameter})}{\Delta d_L(i) } \right]^2
\end{equation}
where $\Delta d_L(i)$ is the observational error bar for each data point indexed by $i$ and $N_{dof}$ is the number of degree of freedom given by the number of data points subtracted to the number of free parameters of the model. We fix $H_0 = 70.4 \frac{\text{km}}{\text{Mpc.s}}$. A comparison of the result for QFC model and the standard flat FLRW model ($\delta_G=\delta_\Lambda = 0$) is presented in the left panel of table \ref{table} together with the 95\% confidence interval. Figure \ref{fig:1} displays the resulting Hubble diagram for both models and the residual of the fit for the QFC model.

From table \ref{table}, we note that the parameters of the Friedmann model are recovered with reasonable precision. However, there is a strong degeneracy between $\delta_G$ and $\Omega_{\rm m}^0$ for the QFC model (the fit gives a correlation coefficient equal to 95\%). To investigate this issue, in section \ref{sec:paramstudy} we inspect in more details the parameter space of the QFC model without imposing any constraints.

Only with SNe Ia the value $\Omega_{\rm m}$ is poorly constrained (see table \ref{table}) but with joined data sets, the current best value is $\Omega_{\rm m}=0.308 \pm 0.012$ \cite{Ade:2015xua}. An important point is that for most of interacting dark energy models both the dark matter and the baryonic matter fluids are assumed to interact with the dark energy, but as the nature of the baryonic matter is known from ground experiment, the type of interaction allowed between this matter and the hypothetical dark energy is strongly constrained. Therefore a more conservative approach would be to allow only dark matter (accounting for $\Omega_{DM} = 0.268 \pm 0.013$ \cite{Ade:2015xua}) to interact with dark energy and account for this interaction with a new term in the bias quantity describing the different behavior of the baryonic matter and the dark matter in the cosmic history. The fact that the best fit from the QFC model is lower ($\Omega_{\rm m} = 0.29$) might be a hint pointing to this requirement of having non-interacting baryonic matter \cite{Amendola:2001rc}. In order to improve the constraint of the parameter $\delta_g$, we now turn to a second dataset and perform a second data analysis in the next section. 

\section{Including data from Baryon Acoustic Oscillations}
In this section, following similar works such as \cite{Sola:2018sjf}, we include data from Baryon Acoustic Oscillation. We use 8 effective BAO points from \cite{Kazin:2014qga,Gil-Marin:2016wya,Gil-Marin:2018cgo,Carter:2018vce}. We compute the QFT prediction for the following quantity:
\begin{equation}
D_V=\left((1+z)^2 D_A^2 \frac{cz}{H(z)} \right)^{1/3},
\end{equation}
where $D_A(z)=\frac{d_L}{(1+z)^2}$ given in \eqref{eq:dl} and $H(z)$ in equation \eqref{re3}.
We minimize the reduced $\chi^2_{BAO}$ defined as:
\begin{equation}
\label{eq:ChiBAO}
\chi^2_{BAO}(\text{parameters}) \equiv \frac{1}{N_{dof}}\sum_{i=1}^{8} \left[ \frac{D_V(i)-D_V(\text{parameter})}{\Delta D_V(i) } \right]^2.
\end{equation}
After minimizing \eqref{eq:ChiBAO}, we perform a joint data analysis of BAO+SNe Ia, minimizing \eqref{eq:chiSN}+\eqref{eq:ChiBAO}. This joint analysis improves the constraint on both $\Omega_{\rm m}$ and $\delta_g$ for the QFC model. Tables \ref{table} sum up our findings.

\begin{table}\centering
\begin{tabular}{|c||c|c|c||c|c|c||c|c|c|}
  \hline
   & $\Omega_{\rm m}^0$ &  $\delta_G$ &$\chi^2_{\text{SNe}}$  & $\Omega_{\rm m}^0$ &  $\delta_G$ &$\chi^2_{\text{BAO}}$ & $\Omega_{\rm m}^0$ &  $\delta_G$ &$\chi^2_{\text{SNe+BAO}}$ \\
  \hline
  flat FLRW & $0.30\pm 0.03$  & - & $0.931$ & $0.30\pm 0.05$  & - & $1.36$& $0.30 \pm 0.02$  & - & $0.934$\\
  QFC model & $0.29 \pm 0.07$ &  $-0.090 \pm 0.7$ & $0.932$& $0.34 \pm 0.22$ &  $0.23 \pm 0.32$ & $1.55$& $0.30 \pm 0.06$ &  $-0.027 \pm 0.38 $ & $0.936$ \\
  \hline
\end{tabular}
\caption{The best fit parameters for the two models under consideration and the associated $\chi^2$, together with the 95 \% confidence intervals. Note that the relation (\ref{eq:rela}) is enforced. The left panel is for supernovae alone, the middle panel for BAO alone and the right panel is for the join fit of supernovae and BAO.}
\label{table}
\end{table}
While the inclusion of another data set helps to reduce the error-bar on the determination of $\Omega_{\rm m}^0$ and $\delta_g$ in the QFC model, we see that they are still a degeneracy between the two parameters of the QFC model and a much more robust data analysis is required in order to break it. Motivated by the results of this section, we now turn on the second scaling exponent $\delta_{\Lambda}$ of the QFC model.
\section{Parameter space study}
\label{sec:paramstudy}
In this section, we drop the relation (\ref{eq:rela}) in order to explore the parameter space $\delta_G$-$\delta_\Lambda$ with supernovae data, for set values of $\Omega_\Lambda^0$ and $\Omega_{\rm m}^0$ compatible with constraints from cosmological observations. The purpose of this section is to identify regions for which the $\chi^2_{\text{SNe}}$ of the QFC model (equation \eqref{eq:chiSN}) is smaller than that of the $\Lambda$CDM model.

\begin{figure}[t]
\centering
\begin{tabular}{ccc}
\includegraphics[width=0.33\textwidth]{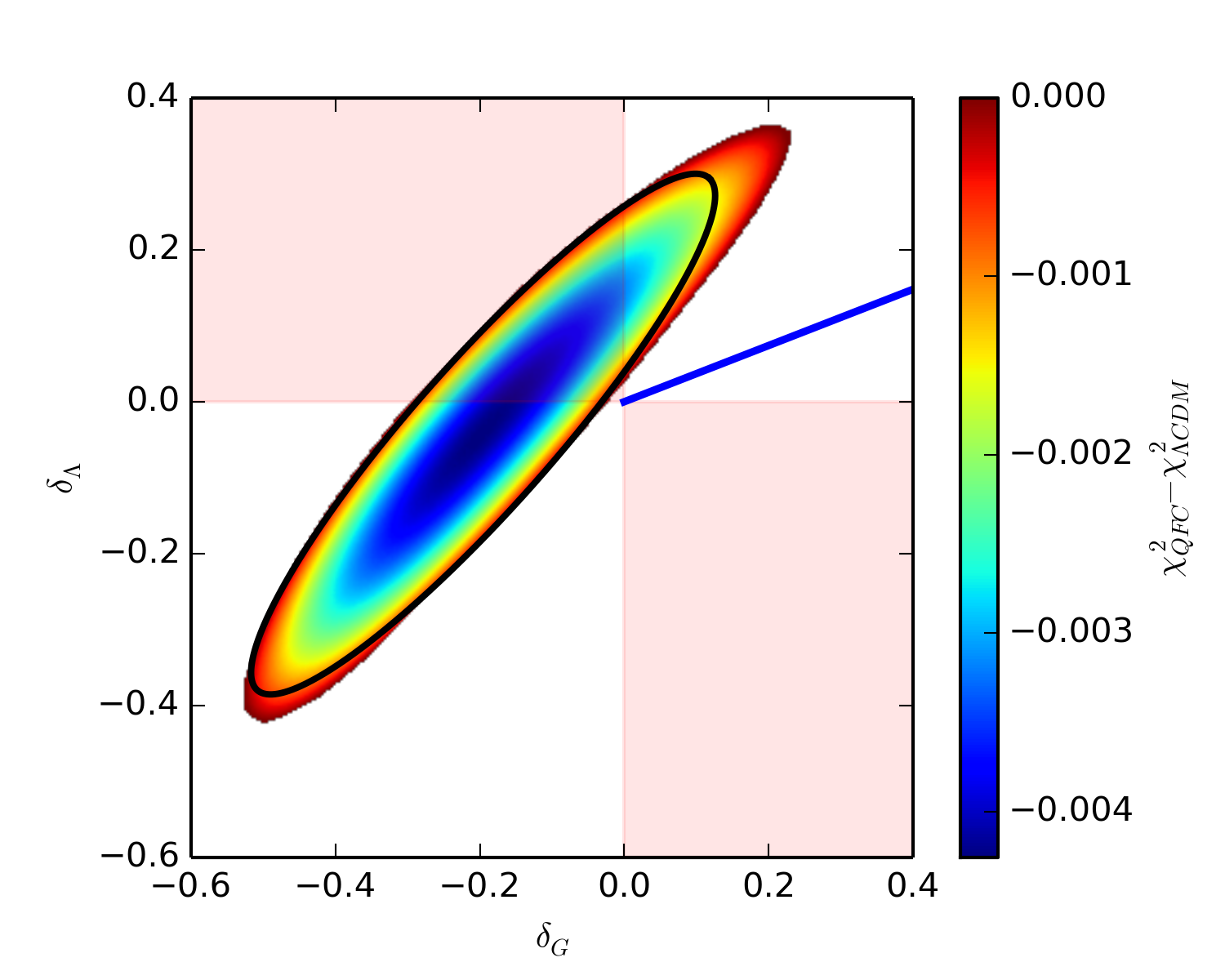} &
\includegraphics[width=0.33\textwidth]{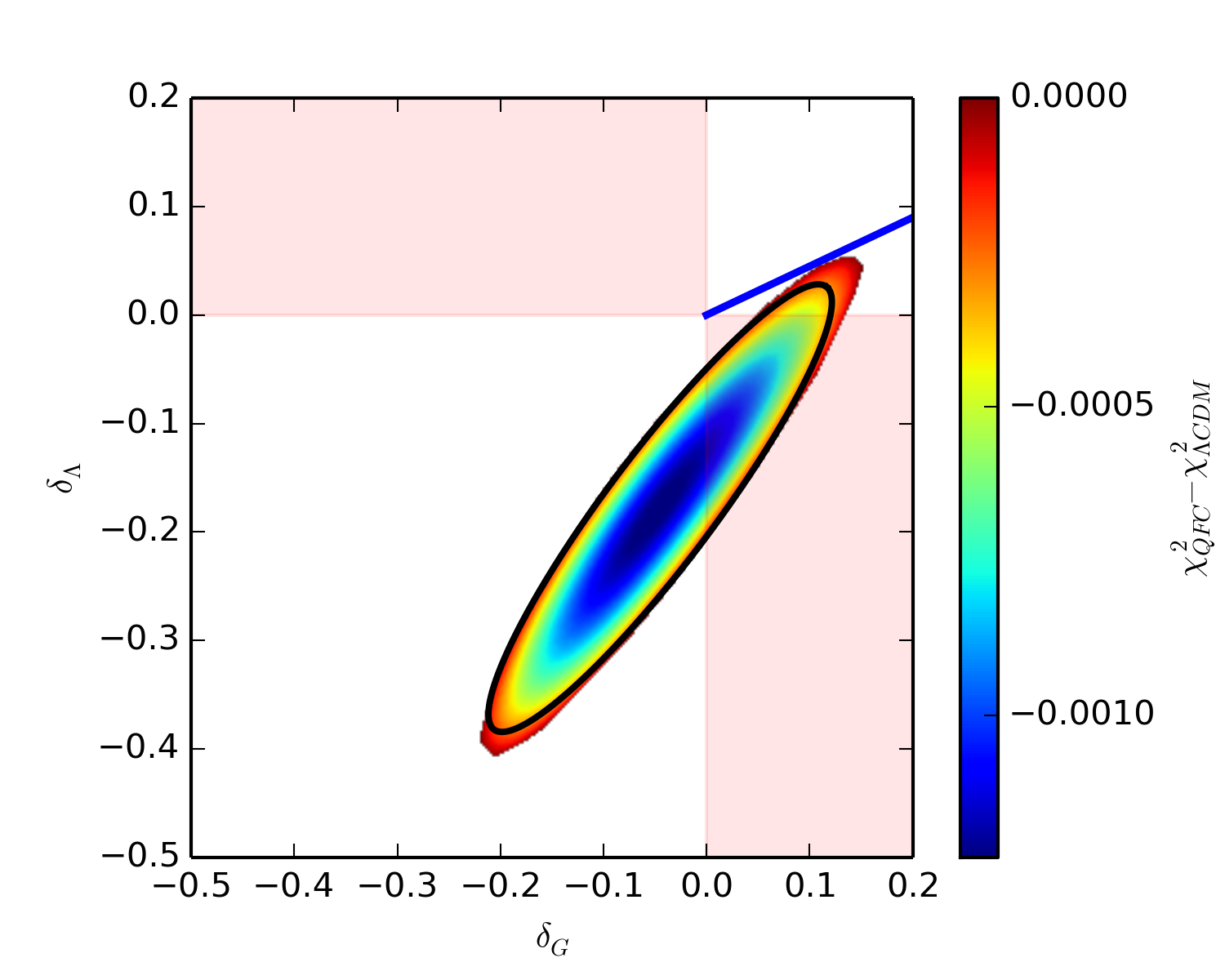} &
\includegraphics[width=0.33\textwidth]{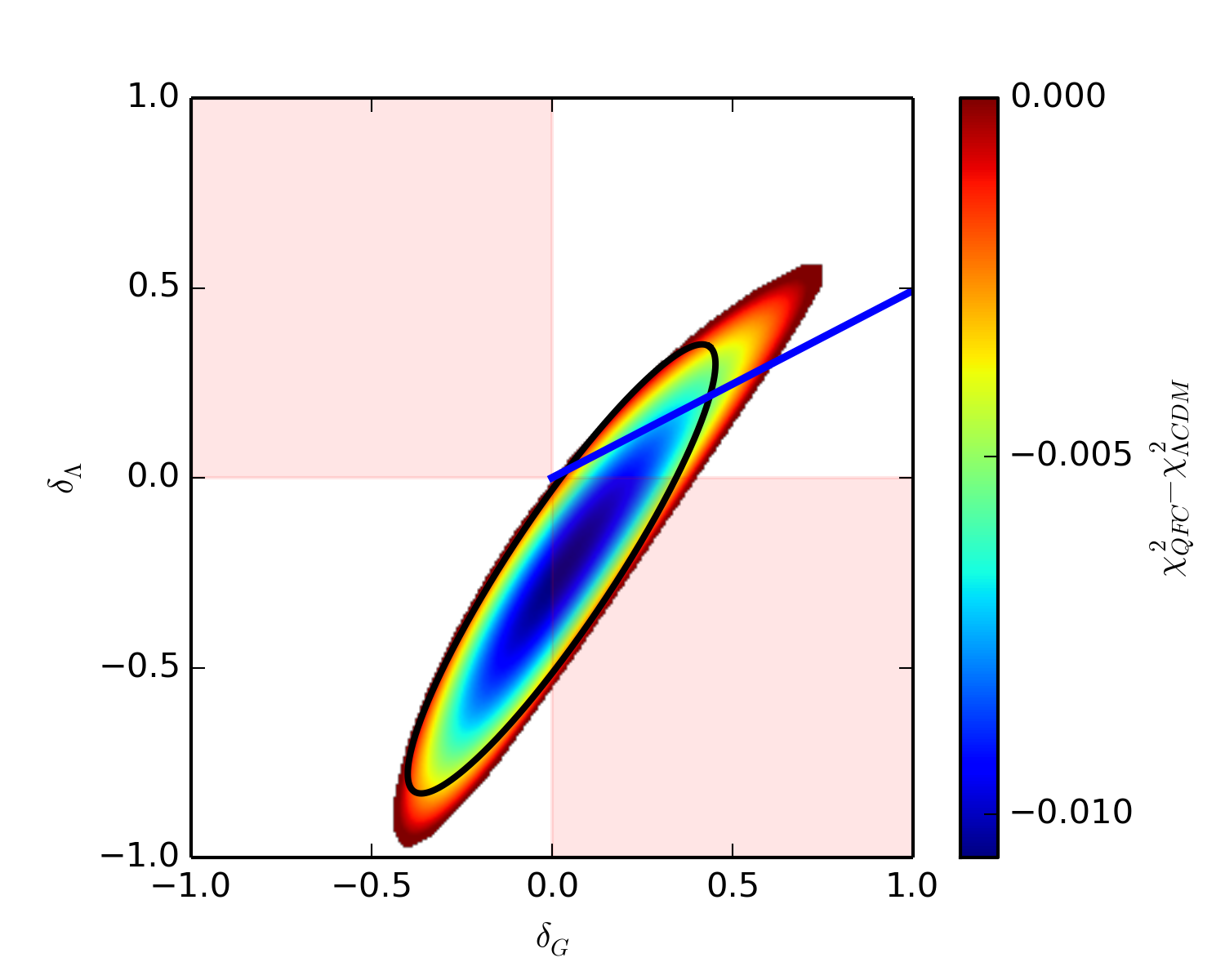}
\end{tabular}
\caption{Contour which obtains a better $\chi^2_{\text{SNe}}$ than the Friedmann model as a function of the two new parameters introduced for different value of $\Omega_{\rm m}^0$. For left to right, we displayed $\Omega_{\rm m}^0=0.27, 0.31, 0.33$, observe the different scale for the difference of $\chi^2_{\text{SNe}}$. The red regions are non-physical with $\delta_G \delta_\Lambda <0$. Equation (\ref{eq:rela}) has been also represented in blue. Observe that for instance for $\Omega_{\rm m}^0=0.33$, it is possible to find values for $\delta_G$ and $\delta_{\Lambda}$ linked by the relation (\ref{eq:rela}) giving a better $\chi^2_{\text{SNe}}$ than the one of the FLRW model.}
\label{fig:3D}
\end{figure}

On figure \ref{fig:3D}, we plotted the difference of $\chi^2_{\text{SNe}}$ between the QFC and Friedmann models for different values of $\Omega_{\rm m}$. It is found that a large zone of the parameter space allows for the QFC model to have a smaller $\chi^2_{\text{SNe}}$ than the one of the Friedmann model, constraining thus the QFC model. Figure \ref{fig:3D} also shows the different quadrants allowing a physical QFC models satisfying the constraint $ \delta_G \delta_{\Lambda} >0$. The linear relation (\ref{eq:rela}) is also displayed in blue. We see that, depending on the value for $\Omega_{\rm m}$, it is not certain to have values for $\delta_G$ and $\delta_{\Lambda}$ allowing for a better $\chi^2_{\text{SNe}}$ than the FLRW together with the relation (\ref{eq:rela}).

\section{Discussion and conclusion}
\label{sec:ccl}
We presented a new model for interacting dark energy arising from the effective field theory of Einstein-Cartan gravity in section \ref{sec:SFC} and explored its observational consequences on supernovae Ia and BAO. We compared and contrasted it with the standard $\Lambda$CDM model assuming equation (\ref{eq:rela}). We then explored the parameter space for this model to know for which value of the parameters, the model has a better $\chi ^2$ than its FLRW counterpart. We found that if one decreases slightly the matter content of the universe, a substantial zone of the parameter space offers a smaller $\chi^2$ than the FLRW model. It could be a possible hint pointing to the need to consider a correction due to interacting dark energy to the halo bias for structure formation, we hope to come back to those questions. It is not the first time that decent alternatives to the $\Lambda$CDM model based on a time variation of the cosmological constant and interaction between dark matter and dark energy are explored. The QFC model provides one more example of such an alternative. 
\par The QFC model of \cite{Xue:2014kna} is also constrained by other independent experiments which constrains the effective variation of the gravitational constant. By Lunar Laser Ranging experiment  \cite{Williams:2004qba}, we find a bound for $\delta_G$:
\begin{equation}
|\delta_G| < 0.02.
\end{equation}
Furthermore the effective variations of the gravitational constant should not spoil the big-bang nucleosynthesis which gives therefore an even stringent constraint on the variation of the gravitational constant which translates for our parameters into:
\begin{equation}
|\delta_G|< \mathcal{O}(10^{-3}),
\end{equation}
depending on the model used for nucleosynthesis \cite{Grande:2010vg}. In our notation, $\delta_g$ corresponds to $\nu$ in \cite{Grande:2010vg}.
\par This study is a first step towards better exploring the parameter space and setting constraints on the parameters. As already done for other interacting dark energy models, it needs to be challenged with different independent data sets. An important extension of this work would be to follow the research plan carried for instance in \cite{Sola:2016ecz} where similar models were confronted to 5 different data sets. We stress that to do so for this model, one needs not only to consider the dynamics explored in equations (\ref{re3}) and (\ref{e2}) which is only valid for small redshift but a more general one given by equation (36) of \cite{Xue:2014kna}. 

\section*{Authors’ contributions}
SSX and CS conceived the study based on the model introduced by SSX. CS calculated the luminosity distance and started the data analysis. The more systematic data analysis in section \ref{sec:paramstudy} was carried out by DB. CS drafted the manuscript. All authors discussed the results, commented on the article and approved the final manuscript.


\begin{thebibliography}{43}
\expandafter\ifx\csname natexlab\endcsname\relax\def\natexlab#1{#1}\fi
\expandafter\ifx\csname bibnamefont\endcsname\relax
  \def\bibnamefont#1{#1}\fi
\expandafter\ifx\csname bibfnamefont\endcsname\relax
  \def\bibfnamefont#1{#1}\fi
\expandafter\ifx\csname citenamefont\endcsname\relax
  \def\citenamefont#1{#1}\fi
\expandafter\ifx\csname url\endcsname\relax
  \def\url#1{\texttt{#1}}\fi
\expandafter\ifx\csname urlprefix\endcsname\relax\def\urlprefix{URL }\fi
\providecommand{\bibinfo}[2]{#2}
\providecommand{\eprint}[2][]{\url{#2}}

\bibitem[{\citenamefont{Riess et~al.}(1998)}]{Riess:1998cb}
\bibinfo{author}{\bibfnamefont{A.~G.} \bibnamefont{Riess}} \bibnamefont{et~al.}
  (\bibinfo{collaboration}{Supernova Search Team}), \bibinfo{journal}{Astron.
  J.} \textbf{\bibinfo{volume}{116}}, \bibinfo{pages}{1009}
  (\bibinfo{year}{1998}), \eprint{astro-ph/9805201}.

\bibitem[{\citenamefont{Perlmutter et~al.}(1999)}]{Perlmutter:1998np}
\bibinfo{author}{\bibfnamefont{S.}~\bibnamefont{Perlmutter}}
  \bibnamefont{et~al.} (\bibinfo{collaboration}{Supernova Cosmology Project}),
  \bibinfo{journal}{Astrophys. J.} \textbf{\bibinfo{volume}{517}},
  \bibinfo{pages}{565} (\bibinfo{year}{1999}), \eprint{astro-ph/9812133}.

\bibitem[{\citenamefont{Ade et~al.}(2016)}]{Ade:2015xua}
\bibinfo{author}{\bibfnamefont{P.~A.~R.} \bibnamefont{Ade}}
  \bibnamefont{et~al.} (\bibinfo{collaboration}{Planck}),
  \bibinfo{journal}{Astron. Astrophys.} \textbf{\bibinfo{volume}{594}},
  \bibinfo{pages}{A13} (\bibinfo{year}{2016}), \eprint{1502.01589}.

\bibitem[{\citenamefont{Tsujikawa}(2013)}]{Tsujikawa:2013fta}
\bibinfo{author}{\bibfnamefont{S.}~\bibnamefont{Tsujikawa}},
  \bibinfo{journal}{Class. Quant. Grav.} \textbf{\bibinfo{volume}{30}},
  \bibinfo{pages}{214003} (\bibinfo{year}{2013}), \eprint{1304.1961}.

\bibitem[{\citenamefont{Buchert}(2008)}]{Buchert:2007ik}
\bibinfo{author}{\bibfnamefont{T.}~\bibnamefont{Buchert}},
  \bibinfo{journal}{Gen. Rel. Grav.} \textbf{\bibinfo{volume}{40}},
  \bibinfo{pages}{467} (\bibinfo{year}{2008}), \eprint{0707.2153}.

\bibitem[{\citenamefont{Bolejko et~al.}(2011)\citenamefont{Bolejko, Celerier,
  and Krasinski}}]{Bolejko:2011jc}
\bibinfo{author}{\bibfnamefont{K.}~\bibnamefont{Bolejko}},
  \bibinfo{author}{\bibfnamefont{M.-N.} \bibnamefont{Celerier}},
  \bibnamefont{and}
  \bibinfo{author}{\bibfnamefont{A.}~\bibnamefont{Krasinski}},
  \bibinfo{journal}{Class. Quant. Grav.} \textbf{\bibinfo{volume}{28}},
  \bibinfo{pages}{164002} (\bibinfo{year}{2011}), \eprint{1102.1449}.

\bibitem[{\citenamefont{Stahl}(2016)}]{Stahl:2016vcl}
\bibinfo{author}{\bibfnamefont{C.}~\bibnamefont{Stahl}}, \bibinfo{journal}{Int.
  J. Mod. Phys.} \textbf{\bibinfo{volume}{D25}}, \bibinfo{pages}{1650066}
  (\bibinfo{year}{2016}).

\bibitem[{\citenamefont{Joyce et~al.}(2015)\citenamefont{Joyce, Jain, Khoury,
  and Trodden}}]{Joyce:2014kja}
\bibinfo{author}{\bibfnamefont{A.}~\bibnamefont{Joyce}},
  \bibinfo{author}{\bibfnamefont{B.}~\bibnamefont{Jain}},
  \bibinfo{author}{\bibfnamefont{J.}~\bibnamefont{Khoury}}, \bibnamefont{and}
  \bibinfo{author}{\bibfnamefont{M.}~\bibnamefont{Trodden}},
  \bibinfo{journal}{Phys. Rept.} \textbf{\bibinfo{volume}{568}},
  \bibinfo{pages}{1} (\bibinfo{year}{2015}), \eprint{1407.0059}.

\bibitem[{\citenamefont{Frieman and Gradwohl}(1991)}]{Friedman:1991dj}
\bibinfo{author}{\bibfnamefont{J.~A.} \bibnamefont{Frieman}} \bibnamefont{and}
  \bibinfo{author}{\bibfnamefont{B.-A.} \bibnamefont{Gradwohl}},
  \bibinfo{journal}{Phys. Rev. Lett.} \textbf{\bibinfo{volume}{67}},
  \bibinfo{pages}{2926} (\bibinfo{year}{1991}).

\bibitem[{\citenamefont{Gradwohl and Frieman}(1992)}]{Gradwohl:1992ue}
\bibinfo{author}{\bibfnamefont{B.-A.} \bibnamefont{Gradwohl}} \bibnamefont{and}
  \bibinfo{author}{\bibfnamefont{J.~A.} \bibnamefont{Frieman}},
  \bibinfo{journal}{Astrophys. J.} \textbf{\bibinfo{volume}{398}},
  \bibinfo{pages}{407} (\bibinfo{year}{1992}).

\bibitem[{\citenamefont{Wetterich}(1995)}]{Wetterich:1994bg}
\bibinfo{author}{\bibfnamefont{C.}~\bibnamefont{Wetterich}},
  \bibinfo{journal}{Astron. Astrophys.} \textbf{\bibinfo{volume}{301}},
  \bibinfo{pages}{321} (\bibinfo{year}{1995}), \eprint{hep-th/9408025}.

\bibitem[{\citenamefont{Amendola}(2000)}]{Amendola:1999er}
\bibinfo{author}{\bibfnamefont{L.}~\bibnamefont{Amendola}},
  \bibinfo{journal}{Phys. Rev.} \textbf{\bibinfo{volume}{D62}},
  \bibinfo{pages}{043511} (\bibinfo{year}{2000}), \eprint{astro-ph/9908023}.

\bibitem[{\citenamefont{Amendola}(2004)}]{Amendola:2003wa}
\bibinfo{author}{\bibfnamefont{L.}~\bibnamefont{Amendola}},
  \bibinfo{journal}{Phys. Rev.} \textbf{\bibinfo{volume}{D69}},
  \bibinfo{pages}{103524} (\bibinfo{year}{2004}), \eprint{astro-ph/0311175}.

\bibitem[{\citenamefont{Lee et~al.}(2006)\citenamefont{Lee, Liu, and
  Ng}}]{Lee:2006za}
\bibinfo{author}{\bibfnamefont{S.}~\bibnamefont{Lee}},
  \bibinfo{author}{\bibfnamefont{G.-C.} \bibnamefont{Liu}}, \bibnamefont{and}
  \bibinfo{author}{\bibfnamefont{K.-W.} \bibnamefont{Ng}},
  \bibinfo{journal}{Phys. Rev.} \textbf{\bibinfo{volume}{D73}},
  \bibinfo{pages}{083516} (\bibinfo{year}{2006}), \eprint{astro-ph/0601333}.

\bibitem[{\citenamefont{Pettorino and Baccigalupi}(2008)}]{Pettorino:2008ez}
\bibinfo{author}{\bibfnamefont{V.}~\bibnamefont{Pettorino}} \bibnamefont{and}
  \bibinfo{author}{\bibfnamefont{C.}~\bibnamefont{Baccigalupi}},
  \bibinfo{journal}{Phys. Rev.} \textbf{\bibinfo{volume}{D77}},
  \bibinfo{pages}{103003} (\bibinfo{year}{2008}), \eprint{0802.1086}.

\bibitem[{\citenamefont{Valiviita et~al.}(2008)\citenamefont{Valiviita,
  Majerotto, and Maartens}}]{Valiviita:2008iv}
\bibinfo{author}{\bibfnamefont{J.}~\bibnamefont{Valiviita}},
  \bibinfo{author}{\bibfnamefont{E.}~\bibnamefont{Majerotto}},
  \bibnamefont{and} \bibinfo{author}{\bibfnamefont{R.}~\bibnamefont{Maartens}},
  \bibinfo{journal}{JCAP} \textbf{\bibinfo{volume}{0807}}, \bibinfo{pages}{020}
  (\bibinfo{year}{2008}), \eprint{0804.0232}.

\bibitem[{\citenamefont{He and Wang}(2008)}]{He:2008tn}
\bibinfo{author}{\bibfnamefont{J.-H.} \bibnamefont{He}} \bibnamefont{and}
  \bibinfo{author}{\bibfnamefont{B.}~\bibnamefont{Wang}},
  \bibinfo{journal}{JCAP} \textbf{\bibinfo{volume}{0806}}, \bibinfo{pages}{010}
  (\bibinfo{year}{2008}), \eprint{0801.4233}.

\bibitem[{\citenamefont{Gavela et~al.}(2009)\citenamefont{Gavela, Hernandez,
  Lopez~Honorez, Mena, and Rigolin}}]{Gavela:2009cy}
\bibinfo{author}{\bibfnamefont{M.~B.} \bibnamefont{Gavela}},
  \bibinfo{author}{\bibfnamefont{D.}~\bibnamefont{Hernandez}},
  \bibinfo{author}{\bibfnamefont{L.}~\bibnamefont{Lopez~Honorez}},
  \bibinfo{author}{\bibfnamefont{O.}~\bibnamefont{Mena}}, \bibnamefont{and}
  \bibinfo{author}{\bibfnamefont{S.}~\bibnamefont{Rigolin}},
  \bibinfo{journal}{JCAP} \textbf{\bibinfo{volume}{0907}}, \bibinfo{pages}{034}
  (\bibinfo{year}{2009}), \bibinfo{note}{[Erratum: JCAP1005,E01(2010)]},
  \eprint{0901.1611}.

\bibitem[{\citenamefont{Faraoni et~al.}(2014)\citenamefont{Faraoni, Dent, and
  Saridakis}}]{Faraoni:2014vra}
\bibinfo{author}{\bibfnamefont{V.}~\bibnamefont{Faraoni}},
  \bibinfo{author}{\bibfnamefont{J.~B.} \bibnamefont{Dent}}, \bibnamefont{and}
  \bibinfo{author}{\bibfnamefont{E.~N.} \bibnamefont{Saridakis}},
  \bibinfo{journal}{Phys. Rev.} \textbf{\bibinfo{volume}{D90}},
  \bibinfo{pages}{063510} (\bibinfo{year}{2014}), \eprint{1405.7288}.

\bibitem[{\citenamefont{Salvatelli et~al.}(2014)\citenamefont{Salvatelli, Said,
  Bruni, Melchiorri, and Wands}}]{Salvatelli:2014zta}
\bibinfo{author}{\bibfnamefont{V.}~\bibnamefont{Salvatelli}},
  \bibinfo{author}{\bibfnamefont{N.}~\bibnamefont{Said}},
  \bibinfo{author}{\bibfnamefont{M.}~\bibnamefont{Bruni}},
  \bibinfo{author}{\bibfnamefont{A.}~\bibnamefont{Melchiorri}},
  \bibnamefont{and} \bibinfo{author}{\bibfnamefont{D.}~\bibnamefont{Wands}},
  \bibinfo{journal}{Phys. Rev. Lett.} \textbf{\bibinfo{volume}{113}},
  \bibinfo{pages}{181301} (\bibinfo{year}{2014}), \eprint{1406.7297}.

\bibitem[{\citenamefont{Xue}(2015{\natexlab{a}})}]{Xue:2014kna}
\bibinfo{author}{\bibfnamefont{S.-S.} \bibnamefont{Xue}},
  \bibinfo{journal}{Nucl. Phys.} \textbf{\bibinfo{volume}{B897}},
  \bibinfo{pages}{326} (\bibinfo{year}{2015}{\natexlab{a}}),
  \eprint{1410.6152}.

\bibitem[{\citenamefont{Ferreira et~al.}(2017)\citenamefont{Ferreira, Quintin,
  Costa, Abdalla, and Wang}}]{Abdalla:2014cla}
\bibinfo{author}{\bibfnamefont{E.~G.~M.} \bibnamefont{Ferreira}},
  \bibinfo{author}{\bibfnamefont{J.}~\bibnamefont{Quintin}},
  \bibinfo{author}{\bibfnamefont{A.~A.} \bibnamefont{Costa}},
  \bibinfo{author}{\bibfnamefont{E.}~\bibnamefont{Abdalla}}, \bibnamefont{and}
  \bibinfo{author}{\bibfnamefont{B.}~\bibnamefont{Wang}},
  \bibinfo{journal}{Phys. Rev.} \textbf{\bibinfo{volume}{D95}},
  \bibinfo{pages}{043520} (\bibinfo{year}{2017}), \eprint{1412.2777}.

\bibitem[{\citenamefont{Koivisto et~al.}(2015)\citenamefont{Koivisto,
  Saridakis, and Tamanini}}]{Koivisto:2015qua}
\bibinfo{author}{\bibfnamefont{T.~S.} \bibnamefont{Koivisto}},
  \bibinfo{author}{\bibfnamefont{E.~N.} \bibnamefont{Saridakis}},
  \bibnamefont{and} \bibinfo{author}{\bibfnamefont{N.}~\bibnamefont{Tamanini}},
  \bibinfo{journal}{JCAP} \textbf{\bibinfo{volume}{1509}}, \bibinfo{pages}{047}
  (\bibinfo{year}{2015}), \eprint{1505.07556}.

\bibitem[{\citenamefont{Kumar and Nunes}(2016)}]{Kumar:2016zpg}
\bibinfo{author}{\bibfnamefont{S.}~\bibnamefont{Kumar}} \bibnamefont{and}
  \bibinfo{author}{\bibfnamefont{R.~C.} \bibnamefont{Nunes}},
  \bibinfo{journal}{Phys. Rev.} \textbf{\bibinfo{volume}{D94}},
  \bibinfo{pages}{123511} (\bibinfo{year}{2016}), \eprint{1608.02454}.

\bibitem[{\citenamefont{Kumar and Nunes}(2017)}]{Kumar:2017dnp}
\bibinfo{author}{\bibfnamefont{S.}~\bibnamefont{Kumar}} \bibnamefont{and}
  \bibinfo{author}{\bibfnamefont{R.~C.} \bibnamefont{Nunes}},
  \bibinfo{journal}{Phys. Rev.} \textbf{\bibinfo{volume}{D96}},
  \bibinfo{pages}{103511} (\bibinfo{year}{2017}), \eprint{1702.02143}.

\bibitem[{\citenamefont{Wang et~al.}(2016)\citenamefont{Wang, Abdalla,
  Atrio-Barandela, and Pavon}}]{Wang:2016lxa}
\bibinfo{author}{\bibfnamefont{B.}~\bibnamefont{Wang}},
  \bibinfo{author}{\bibfnamefont{E.}~\bibnamefont{Abdalla}},
  \bibinfo{author}{\bibfnamefont{F.}~\bibnamefont{Atrio-Barandela}},
  \bibnamefont{and} \bibinfo{author}{\bibfnamefont{D.}~\bibnamefont{Pavon}},
  \bibinfo{journal}{Rept. Prog. Phys.} \textbf{\bibinfo{volume}{79}},
  \bibinfo{pages}{096901} (\bibinfo{year}{2016}), \eprint{1603.08299}.

\bibitem[{\citenamefont{Murgia et~al.}(2016)\citenamefont{Murgia, Gariazzo, and
  Fornengo}}]{Murgia:2016ccp}
\bibinfo{author}{\bibfnamefont{R.}~\bibnamefont{Murgia}},
  \bibinfo{author}{\bibfnamefont{S.}~\bibnamefont{Gariazzo}}, \bibnamefont{and}
  \bibinfo{author}{\bibfnamefont{N.}~\bibnamefont{Fornengo}},
  \bibinfo{journal}{JCAP} \textbf{\bibinfo{volume}{1604}}, \bibinfo{pages}{014}
  (\bibinfo{year}{2016}), \eprint{1602.01765}.

\bibitem[{\citenamefont{Koyama et~al.}(2009)\citenamefont{Koyama, Maartens, and
  Song}}]{Koyama:2009gd}
\bibinfo{author}{\bibfnamefont{K.}~\bibnamefont{Koyama}},
  \bibinfo{author}{\bibfnamefont{R.}~\bibnamefont{Maartens}}, \bibnamefont{and}
  \bibinfo{author}{\bibfnamefont{Y.-S.} \bibnamefont{Song}},
  \bibinfo{journal}{JCAP} \textbf{\bibinfo{volume}{0910}}, \bibinfo{pages}{017}
  (\bibinfo{year}{2009}), \eprint{0907.2126}.

\bibitem[{\citenamefont{Xue}(2015{\natexlab{b}})}]{Xue:2015tmw}
\bibinfo{author}{\bibfnamefont{S.-S.} \bibnamefont{Xue}},
  \bibinfo{journal}{Int. J. Mod. Phys.} \textbf{\bibinfo{volume}{A30}},
  \bibinfo{pages}{1545003} (\bibinfo{year}{2015}{\natexlab{b}}).

\bibitem[{\citenamefont{Solà et~al.}(2017)\citenamefont{Solà, Gómez-Valent,
  and de~Cruz~Pérez}}]{Sola:2016jky}
\bibinfo{author}{\bibfnamefont{J.}~\bibnamefont{Solà}},
  \bibinfo{author}{\bibfnamefont{A.}~\bibnamefont{Gómez-Valent}},
  \bibnamefont{and}
  \bibinfo{author}{\bibfnamefont{J.}~\bibnamefont{de~Cruz~Pérez}},
  \bibinfo{journal}{Astrophys. J.} \textbf{\bibinfo{volume}{836}},
  \bibinfo{pages}{43} (\bibinfo{year}{2017}), \eprint{1602.02103}.
\bibitem[{\citenamefont{Novikov}(2016{\natexlab{a}})}]{Novikov:2016hrc}
\bibinfo{author}{\bibfnamefont{E.~A.} \bibnamefont{Novikov}},
  \bibinfo{journal}{Mod. Phys. Lett.} \textbf{\bibinfo{volume}{A31}},
  \bibinfo{pages}{1650092} (\bibinfo{year}{2016}{\natexlab{a}}).
\bibitem[{\citenamefont{Novikov}(2016{\natexlab{b}})}]{Novikov:2016fzd}
\bibinfo{author}{\bibfnamefont{E.~A.} \bibnamefont{Novikov}},
  \bibinfo{journal}{Electron. J. Theor. Phys.} \textbf{\bibinfo{volume}{13}},
  \bibinfo{pages}{79} (\bibinfo{year}{2016}{\natexlab{b}}).
\bibitem[{\citenamefont{Weinberg}(1977)}]{Weinberg}
\bibinfo{author}{\bibfnamefont{S.}~\bibnamefont{Weinberg}},
  \bibinfo{journal}{Plenum Press., New York}  (\bibinfo{year}{1977}).

\bibitem[{\citenamefont{Suzuki et~al.}(2012)}]{Union}
\bibinfo{author}{\bibfnamefont{N.}~\bibnamefont{Suzuki}} \bibnamefont{et~al.},
  \bibinfo{journal}{Astrophys. J.} \textbf{\bibinfo{volume}{746}},
  \bibinfo{pages}{85} (\bibinfo{year}{2012}), \eprint{1105.3470}.

\bibitem[{\citenamefont{Amendola and
  Tocchini-Valentini}(2002)}]{Amendola:2001rc}
\bibinfo{author}{\bibfnamefont{L.}~\bibnamefont{Amendola}} \bibnamefont{and}
  \bibinfo{author}{\bibfnamefont{D.}~\bibnamefont{Tocchini-Valentini}},
  \bibinfo{journal}{Phys. Rev.} \textbf{\bibinfo{volume}{D66}},
  \bibinfo{pages}{043528} (\bibinfo{year}{2002}), \eprint{astro-ph/0111535}.
\bibitem[{\citenamefont{Peracaula et~al.}(2018)\citenamefont{Peracaula,
  Gomez-Valent, and Perez}}]{Sola:2018sjf}
\bibinfo{author}{\bibfnamefont{J.~S.} \bibnamefont{Peracaula}},
  \bibinfo{author}{\bibfnamefont{A.}~\bibnamefont{Gomez-Valent}},
  \bibnamefont{and} \bibinfo{author}{\bibfnamefont{J.~d.~C.}
  \bibnamefont{Perez}} (\bibinfo{year}{2018}), \eprint{1811.03505}.
\bibitem[{\citenamefont{Kazin et~al.}(2014)}]{Kazin:2014qga}
\bibinfo{author}{\bibfnamefont{E.~A.} \bibnamefont{Kazin}}
  \bibnamefont{et~al.}, \bibinfo{journal}{Mon. Not. Roy. Astron. Soc.}
  \textbf{\bibinfo{volume}{441}}, \bibinfo{pages}{3524} (\bibinfo{year}{2014}),
  \eprint{1401.0358}.
\bibitem[{\citenamefont{Gil-Marín et~al.}(2017)\citenamefont{Gil-Marín,
  Percival, Verde, Brownstein, Chuang, Kitaura, Rodríguez-Torres, and
  Olmstead}}]{Gil-Marin:2016wya}
\bibinfo{author}{\bibfnamefont{H.}~\bibnamefont{Gil-Marín}},
  \bibinfo{author}{\bibfnamefont{W.~J.} \bibnamefont{Percival}},
  \bibinfo{author}{\bibfnamefont{L.}~\bibnamefont{Verde}},
  \bibinfo{author}{\bibfnamefont{J.~R.} \bibnamefont{Brownstein}},
  \bibinfo{author}{\bibfnamefont{C.-H.} \bibnamefont{Chuang}},
  \bibinfo{author}{\bibfnamefont{F.-S.} \bibnamefont{Kitaura}},
  \bibinfo{author}{\bibfnamefont{S.~A.} \bibnamefont{Rodríguez-Torres}},
  \bibnamefont{and} \bibinfo{author}{\bibfnamefont{M.~D.}
  \bibnamefont{Olmstead}}, \bibinfo{journal}{Mon. Not. Roy. Astron. Soc.}
  \textbf{\bibinfo{volume}{465}}, \bibinfo{pages}{1757} (\bibinfo{year}{2017}),
  \eprint{1606.00439}.
\bibitem[{\citenamefont{Gil-Marín et~al.}(2018)}]{Gil-Marin:2018cgo}
\bibinfo{author}{\bibfnamefont{H.}~\bibnamefont{Gil-Marín}}
  \bibnamefont{et~al.}, \bibinfo{journal}{Mon. Not. Roy. Astron. Soc.}
  \textbf{\bibinfo{volume}{477}}, \bibinfo{pages}{1604} (\bibinfo{year}{2018}),
  \eprint{1801.02689}.
\bibitem[{\citenamefont{Carter et~al.}(2018)\citenamefont{Carter, Beutler,
  Percival, Blake, Koda, and Ross}}]{Carter:2018vce}
\bibinfo{author}{\bibfnamefont{P.}~\bibnamefont{Carter}},
  \bibinfo{author}{\bibfnamefont{F.}~\bibnamefont{Beutler}},
  \bibinfo{author}{\bibfnamefont{W.~J.} \bibnamefont{Percival}},
  \bibinfo{author}{\bibfnamefont{C.}~\bibnamefont{Blake}},
  \bibinfo{author}{\bibfnamefont{J.}~\bibnamefont{Koda}}, \bibnamefont{and}
  \bibinfo{author}{\bibfnamefont{A.~J.} \bibnamefont{Ross}}
  (\bibinfo{year}{2018}), \eprint{1803.01746}.
\bibitem[{\citenamefont{Williams et~al.}(2004)\citenamefont{Williams, Turyshev,
  and Boggs}}]{Williams:2004qba}
\bibinfo{author}{\bibfnamefont{J.~G.} \bibnamefont{Williams}},
  \bibinfo{author}{\bibfnamefont{S.~G.} \bibnamefont{Turyshev}},
  \bibnamefont{and} \bibinfo{author}{\bibfnamefont{D.~H.} \bibnamefont{Boggs}},
  \bibinfo{journal}{Phys. Rev. Lett.} \textbf{\bibinfo{volume}{93}},
  \bibinfo{pages}{261101} (\bibinfo{year}{2004}), \eprint{gr-qc/0411113}.

\bibitem[{\citenamefont{Grande et~al.}(2010)\citenamefont{Grande, Sola, Fabris,
  and Shapiro}}]{Grande:2010vg}
\bibinfo{author}{\bibfnamefont{J.}~\bibnamefont{Grande}},
  \bibinfo{author}{\bibfnamefont{J.}~\bibnamefont{Sola}},
  \bibinfo{author}{\bibfnamefont{J.~C.} \bibnamefont{Fabris}},
  \bibnamefont{and} \bibinfo{author}{\bibfnamefont{I.~L.}
  \bibnamefont{Shapiro}}, \bibinfo{journal}{Class. Quant. Grav.}
  \textbf{\bibinfo{volume}{27}}, \bibinfo{pages}{105004}
  (\bibinfo{year}{2010}), \eprint{1001.0259}.
\bibitem[{\citenamefont{Solà~Peracaula
  et~al.}(2018)\citenamefont{Solà~Peracaula, de~Cruz~Pérez, and
  Gómez-Valent}}]{Sola:2016ecz}
\bibinfo{author}{\bibfnamefont{J.}~\bibnamefont{Solà~Peracaula}},
  \bibinfo{author}{\bibfnamefont{J.}~\bibnamefont{de~Cruz~Pérez}},
  \bibnamefont{and}
  \bibinfo{author}{\bibfnamefont{A.}~\bibnamefont{Gómez-Valent}},
  \bibinfo{journal}{EPL} \textbf{\bibinfo{volume}{121}}, \bibinfo{pages}{39001}
  (\bibinfo{year}{2018}), \eprint{1606.00450}.
\end{thebibliography}
\end{document}